\documentclass{llncs}
\usepackage[pdftex]{graphicx}

\usepackage{caption}
\usepackage{mathrsfs}

\usepackage{xspace}

\title{Radiation hardness of small-pitch 3D pixel sensors up to HL-LHC fluences}

\author{J.~Lange\inst{1} \and S.~Grinstein\inst{1,2} \and M.~Manna\inst{3} \and G.~Pellegrini\inst{3} \and D.~Quirion\inst{3} \and S.~Terzo\inst{1} \and D.~V\'{a}zquez Furelos\inst{1}}

\institute{Institut de F\'{i}sica d'Altes Energies (IFAE), The Barcelona Institute of Science and Technology (BIST), 08193 Bellaterra (Barcelona), Spain
\and
Instituci\'{o} Catalana de Recerca i Estudis Avan\c{c}ats (ICREA), Pg. Llu\'{i}s Companys 23, 08010 Barcelona, Spain
\and
Centro Nacional de Microelectronica (CNM-IMB-CSIC), Campus UAB, 08193 Bellaterra (Barcelona), Spain
}

\begin{document}

\maketitle 
\flushbottom 

\begin{abstract}

A new generation of 3D silicon pixel detectors with a small pixel size of 50$\times$50 and 25$\times$100\,$\mu$m$^{2}$ is being developed for the HL-LHC tracker upgrades. The radiation hardness of such detectors was studied in beam tests after irradiation to HL-LHC fluences up to $1.4\times10^{16}$\,n$_{\mathrm{eq}}$/cm$^2$. At this fluence, an operation voltage of only 100\,V is needed to achieve 97\% hit efficiency, with a power dissipation of 13\,mW/cm$^2$ at -25$^{\circ}$C, considerably lower than for previous 3D sensor generations and planar sensors.

\end{abstract}

\section{Introduction}
\label{sec:intro}

3D silicon detectors are studied as radiation-hard candidates for the upgrade of the innermost pixel layers of the ATLAS experiment at the High-Luminosity (HL)-LHC~\cite{bib:HL-LHC}. The detectors are required to withstand unprecedented radiation levels with 1\,MeV neutron equivalent fluences up to 1--2$\times10^{16}$\,n$_{\mathrm{eq}}$/cm$^2$ and increased occupancies.

3D detectors have columnar electrodes inside the sensor bulk instead of at the surface like for planar sensors, so that the electrode distance can be chosen to be significantly smaller than the thickness~\cite{bib:3D}. This gives advantages in terms of less trapping at radiation-induced defects and lower operational voltages, which translates into lower power dissipation after heavy irradiation. The current generation of 3D pixel detectors installed in the ATLAS Insertable B-Layer (IBL)~\cite{bib:IBLprototypes} and the ATLAS Forward Proton (AFP) detector~\cite{bib:AFPproduction} consists of 50$\times$250\,$\mu$m$^{2}$ pixels with two 3D electrodes each (2E), which are connected to the FE-I4 readout chip~\cite{bib:FEI4}. Their 3D inter-electrode distance is $L_{el}=$67\,$\mu$m at a sensor thickness of 230\,$\mu$m. This generation demonstrated already a good radiation hardness up to at least $9\times10^{15}$\,n$_{\mathrm{eq}}$/cm$^2$~\cite{bib:CNMIBLgenAndSmallPitch2}. For the HL-LHC, a new generation of 3D detectors is developed with smaller pixel sizes of 50$\times$50 or 25$\times$100\,$\mu$m$^{2}$, originally motivated by occupancy reasons. This also leads to a significant reduction of $L_{el}$ down to e.g. 35\,$\mu$m for 50$\times$50\,$\mu$m$^{2}$ with one 3D electrode per pixel (1E) or 27\,$\mu$m for 25$\times$100\,$\mu$m$^{2}$ 2E, thereby further improving the radiation hardness.

In this paper, the radiation hardness of a first prototype production of small-pitch 3D detectors by CNM (Centro Nacional de Microelectronica) in Barcelona is studied up to HL-LHC fluences for the first time.


\section{Small-Pitch 3D Pixel Devices}
\label{sec:Devices}

At CNM, a first 3D sensor run (7781) with small 3D cell sizes was carried out on 230\,$\mu$m thick wafers in the CNM double-sided process technology~\cite{bib:CNMIBLProduction}. Since a new readout chip with small pixel size is still under development, the small 3D sensor pixels are matched to the existing FE-I4 readout chip: each 50$\times$250\,$\mu$m$^{2}$ FE-I4 chip pixel cell contains five 50$\times$50\,$\mu$m$^{2}$ 1E or 25$\times$100\,$\mu$m$^{2}$ 2E sensor pixels, so that only 20\% of the sensor pixels can be connected to a read-out channel. The remaining 80\% insensitive sensor pixels are shorted to ground. The pixel sensors were bump-bonded to the FE-I4 readout chip, assembled on readout boards and initially characterised at IFAE Barcelona. More details of the run and initial laboratory and beam-test characterisations are described in Refs.~\cite{bib:CNMsmallPitch1,bib:CNMIBLgenAndSmallPitch2}.

\section{Irradiations}
\label{sec:Irradiations}

Two complementary irradiation campaigns have been carried out at the Karlsruhe Institute of Technology (KIT) and CERN-PS IRRAD facilities.

Two 50$\times$50\,$\mu$m$^{2}$ pixel devices (7781-W3-C1 and 7781-W5-C2) have been irradiated uniformly to 5$\times10^{15}$\,n$_{\mathrm{eq}}$/cm$^2$ at KIT with 23\,MeV protons. This irradiation and fluence is the same as used in the IBL qualification campaigns and hence allows a direct comparison to the IBL generation. The fluence uniformity also allows a simple determination of the power dissipation. However, the low-energy protons are delivering a high ionising dose to the FE-I4 readout chip and hence higher fluences are likely to damage the chip.

Thus, to reach higher fluences relevant for the HL-LHC upgrade, irradiations have been carried out at the CERN-PS IRRAD facility with 24\,GeV protons, which provides a non-uniform fluence over the detector area. Hence, it was possible to study a broad range of fluences on one single pixel detector as already demonstrated in Ref.~\cite{bib:CNMIBLgenAndSmallPitch2}. The Gaussian beam profile has been determined with beam position monitors to 20.4$\times$18.3\,mm$^2$ FWHM, and the overall normalisation has been obtained with gamma spectroscopy of a 20x20\,mm$^{2}$ aluminium foil, which gave an average fluence of 1.0$\times10^{16}$\,n$_{\mathrm{eq}}$/cm$^2$ over this area. The fluence in the studied region ranged from 0.8 to 1.4$\times10^{16}$\,n$_{\mathrm{eq}}$/cm$^2$. Systematic fluence uncertainties have been estimated between 11\% (high fluence) and 24\% (low fluence) using beam centre, beam width and aluminium position variations by 1\,mm. It should be noted that from the initially three irradiated devices (two 50$\times$50 and one 25$\times$100\,$\mu$m$^{2}$), eventually only one device was operational (7781-W4-C1, 50$\times$50\,$\mu$m$^{2}$) with the other two being non-responsive due to either readout chip or board failure.

All devices were annealed for one week at room temperature.

\section{Performance in Beam Tests}
\label{sec:beamTests}

In 2016, several beam tests were performed at the CERN SPS H6 beam line with 120\,GeV pions. A EUDET-type telescope made of six MIMOSA planes~\cite{bib:EUDET} was used to provide reference tracks for hit efficiency determination in the selected active region.

Before irradiation, the small-pitch 3D devices showed a good performance of 96--97\% efficiency at 0$^{\circ}$ tilt already from 1\,V on, indicating a very low full depletion voltage due to the small 3D electrode distance~\cite{bib:CNMIBLgenAndSmallPitch2}.

The irradiated devices were cooled in a climate chamber with set temperatures between -40 and -50$^{\circ}$C, corresponding to on-sensor temperatures between -25 and -35$^{\circ}$C. No influence of temperature on the performance was found in this range. The devices were tuned to a threshold of 1.0 and 1.5\,ke$^{-}$. Measurements were performed at 0$^{\circ}$ and 15$^{\circ}$ tilt with respect to the beam axis.

\begin{figure}[hbt]
	\centering
	\includegraphics[width=8cm]{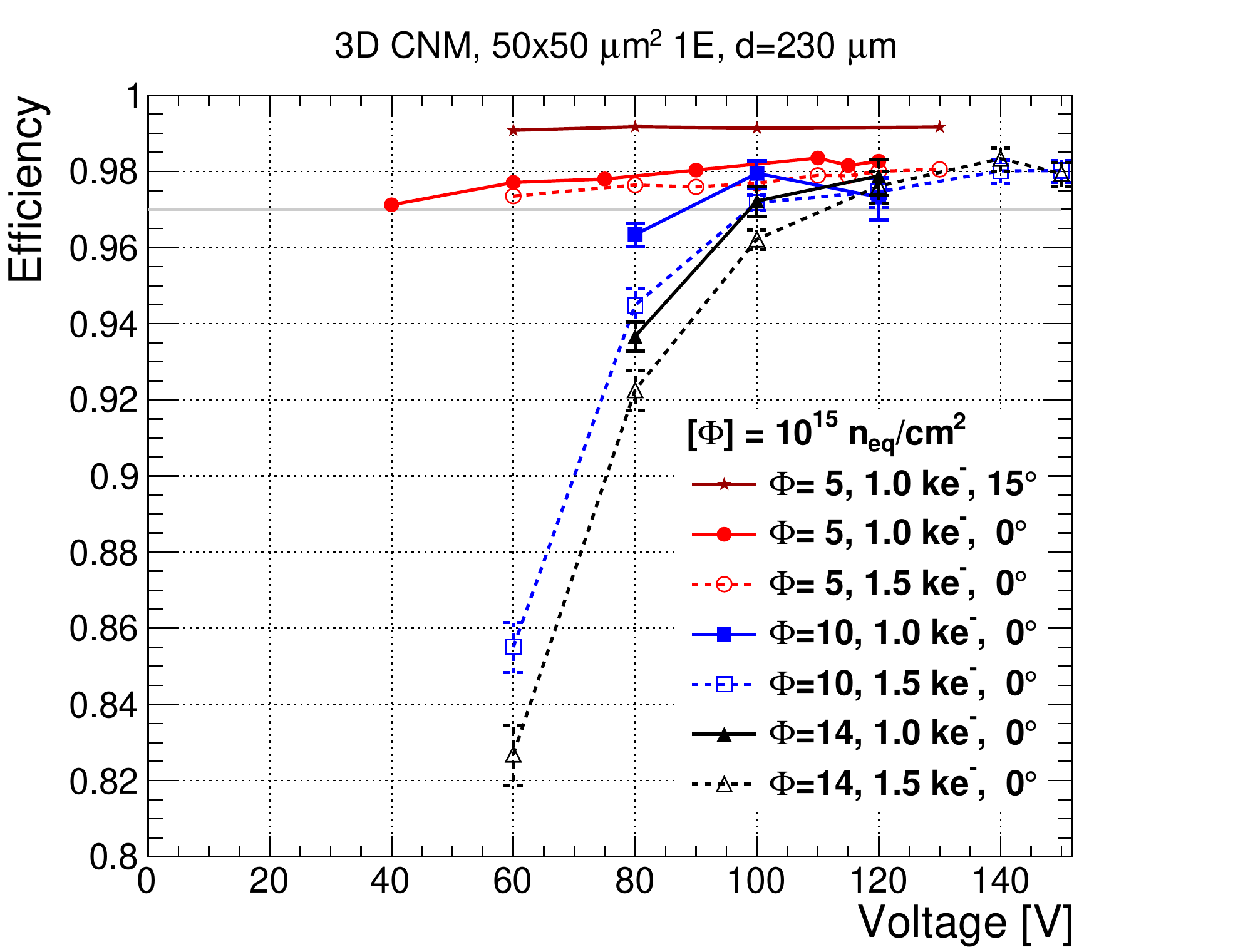}
	\caption{Efficiency as a function of voltage for different fluences, thresholds and tilts. The uncertainties are statistical only.}
	\label{fig:Eff}
\end{figure}

Figure~\ref{fig:Eff} shows the hit efficiency as a function of bias voltage for different fluences, thresholds and tilts for 7781-W3-C1 (5$\times10^{15}$\,n$_{\mathrm{eq}}$/cm$^2$) and 7781-W4-C1 (1.0 and 1.4$\times10^{16}$\,n$_{\mathrm{eq}}$/cm$^2$). As expected, the efficiency is higher for lower fluences, lower thresholds and 15$^{\circ}$ tilt (since situations in which the particle passes exactly through the insensitive 3D columns are avoided). At 5$\times10^{15}$\,n$_{\mathrm{eq}}$/cm$^2$ and 0$^{\circ}$, the benchmark efficiency of 97\% is already reached at $V_{97\%}$=40\,V. Tilting the device by 15$^{\circ}$ increases the efficiency above 99\%. At the highest fluence measured of 1.4$\times10^{16}$\,n$_{\mathrm{eq}}$/cm$^2$, the efficiency is surpassing 97\% at 100\,V for 1.0\,ke$^{-}$ threshold and 0$^{\circ}$ tilt.

\begin{figure}[hbt]
	\centering
	 \includegraphics[width=8cm]{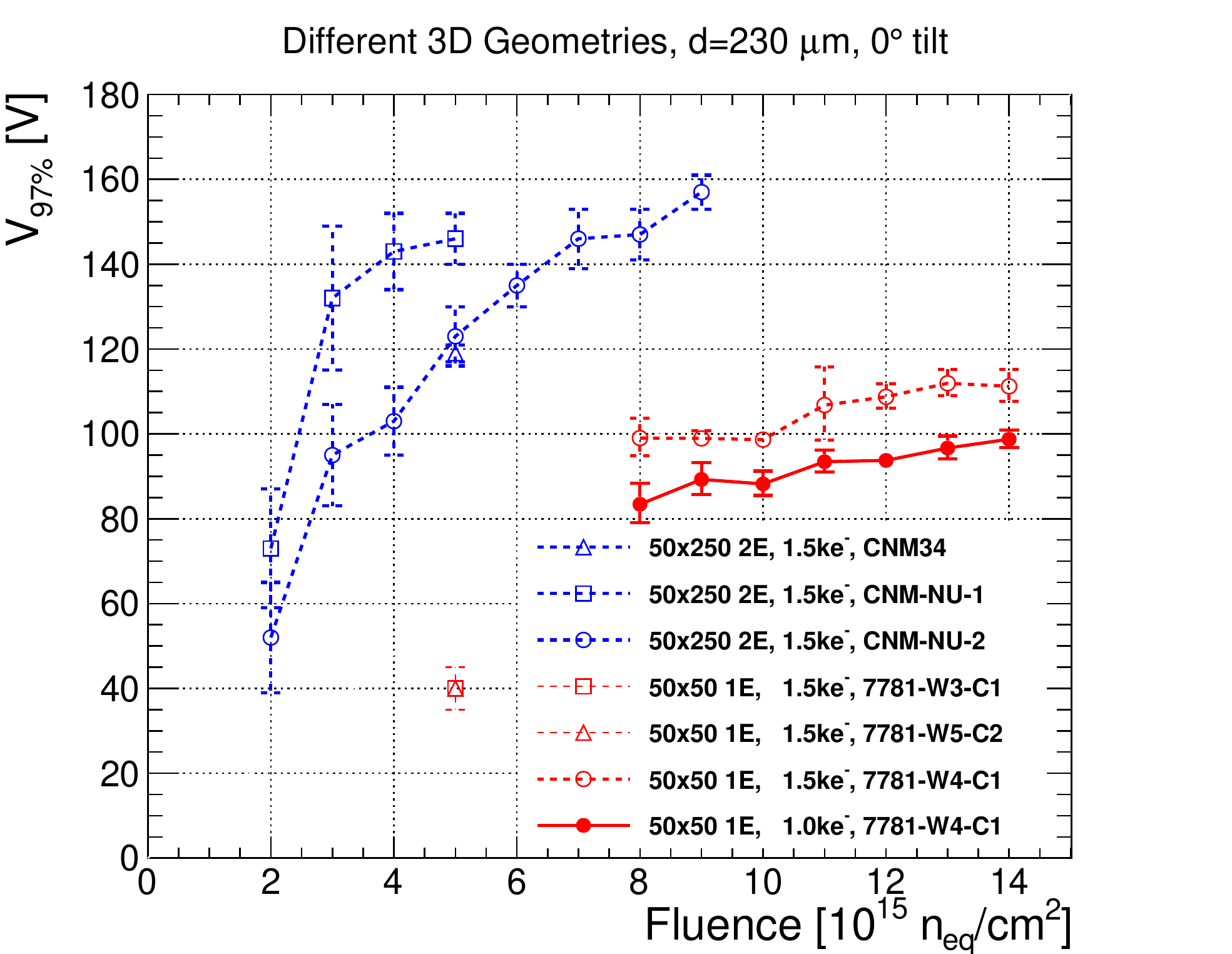}
	\caption{$V_{97\%}$ as a function of fluence, compared for the IBL generation (50x250\,$\mu$m$^2$ 2E) and the new 50x50\,$\mu$m$^2$ 1E pixels.}
	\label{fig:Vop}
\end{figure}

The voltage $V_{97\%}$ to reach the benchmark efficiency of 97\% is taken as an indication for a suitable operation voltage. It is obtained from linear interpolation and shown in Fig.~\ref{fig:Vop} as a function of fluence, compared for the new small-pitch and the IBL 3D generations~\cite{bib:CNMIBLgenAndSmallPitch2}. The significant improvement due to the smaller 3D electrode distance and hence less trapping is visible. For example at 5$\times10^{15}$\,n$_{\mathrm{eq}}$/cm$^2$, the IBL-generation devices need 120\,V to reach 97\% efficiency for 0$^{\circ}$ tilt and 1.5\,ke$^{-}$ threshold, compared to only 40\,V for a 3D cell of 50x50\,$\mu$m$^2$ 1E under the same conditions. This results also in a significant reduction of power dissipation (product of $V_{97\%}$ and leakage current at -25$^{\circ}$C) from 3.5\,mW/cm$^2$ for IBL to 1.5\,mW/cm$^2$ for 50x50\,$\mu$m$^2$. For the non-uniformly irradiated device, the estimation of power dissipation cannot be directly obtained, but is estimated by combining $V_{97\%}$ with the leakage current measured on strip test structures with the same 3D cell size and fluence~\cite{bib:CNMsmallPitch1} as 9 and 13 mW/cm$^{2}$ at 1.0 and 1.4$\times10^{16}$\,n$_{\mathrm{eq}}$/cm$^2$, respectively. Hence, the operation voltage and power dissipation are considerably lower than for planar pixels, for which the best values at 1.0$\times10^{16}$\,n$_{\mathrm{eq}}$/cm$^2$ (obtained for 100\,$\mu$m thickness) are $V_{97\%}$=500\,V with a power dissipation of minimally 25\,mW/cm$^{2}$~\cite{bib:planarPower}.


\section{Conclusions and outlook}
\label{sec:conclusions}

3D silicon pixel detectors have been investigated as candidates for the innermost pixel layers for the HL-LHC upgrade of the ATLAS detector. 
Small-pitch prototypes of 230\,$\mu$m thickness have been produced and studied in beam tests after irradiation to HL-LHC fluences. 

At $1.4\times10^{16}$\,n$_{\mathrm{eq}}$/cm$^2$, an operation voltage of only 100\,V is needed to achieve 97\% hit efficiency, with a power dissipation of 13\,mW/cm$^2$ at -25$^{\circ}$C, considerably lower than for previous 3D generations and planar sensors. Hence, an excellent radiation hardness of small-pitch 3D sensors was found at HL-LHC fluences.

Further irradiation studies of these devices up to even higher fluences above $2\times10^{16}$\,n$_{\mathrm{eq}}$/cm$^2$ are being carried out.

In parallel, further 3D productions are on-going at CNM with improved processes in order to increase the yield and breakdown voltages, as has been already demonstrated~\cite{bib:AFPproduction}. The new runs include productions on thinner wafers (70--150\,$\mu$m) as well as 3D sensors compatible with the new small-pitch RD53A readout chip, besides the FE-I4 prototypes.

\section*{Acknowledgments}
The authors wish to thank A.\,Rummler, M.\,Bomben and the other ATLAS ITk beam test participants for great support and discussions at the beam tests; also to F.\,Ravotti and G.\,Pezzullo (CERN IRRAD) and F.\,B\"{o}gelspacher for excellent support for the irradiations. This work was partly performed in the framework of the CERN RD50 collaboration.This work was partially funded by: the MINECO, Spanish Government, under grants FPA2013-48308-C2-1-P, FPA2015-69260-C3-2-R, FPA2015-69260-C3-3-R (co-financed with the European Union's FEDER funds) and SEV-2012-0234 (Severo Ochoa excellence programme) and under the Juan de la Cierva programme; the Catalan Government (AGAUR): Grups de Recerca Consolidats (SGR 2014 1177); and the European Union's Horizon 2020 Research and Innovation programme under Grant Agreement no. 654168 (AIDA2020).



\providecommand{\href}[2]{#2}\begingroup\raggedright\endgroup

\end{document}